\documentclass[conference]{IEEEtran}
\IEEEoverridecommandlockouts
\pdfoutput=1
\usepackage{calc}
\usepackage{amsmath,amssymb,amsfonts}
\usepackage{algorithm}
\usepackage{algpseudocode}
\usepackage{graphicx}
\usepackage{textcomp}
\usepackage{xcolor}
\usepackage{listings}
\usepackage{booktabs}
\usepackage[group-separator={,}, group-minimum-digits={3}]{siunitx}
\usepackage[colorlinks=true, urlcolor=blue,
            linkcolor=blue, anchorcolor=blue]{hyperref}
\usepackage{paralist}
\usepackage[expansion,protrusion]{microtype}
\usepackage{todonotes}
\usepackage{numprint}
\usepackage{tikz}
\usepackage{cite}
\usetikzlibrary{calc, positioning, shapes.geometric}

\newcommand{\nop}[1]{{}} 

\newcommand{\simulationLoC}{114, 527, 601}
\newcommand{\optimizationLoC}{64, 176, 345, 458}
\newcommand{\mlLoC}{170, 206, 248, 249, 1322, 1794}
\newcommand{\averageLoC}{434}

\newcommand{\LoCcolumnwidth}{90} 
\newcommand{\LoCunit}{\LoCcolumnwidth/1800}
\newcommand{\dotposition}[1]{\LoCunit*#1}

\newcommand{\random}{rand*1 pt}

\tikzset{line/.style={very_lightgray,very thin}}
\newcommand{\LoCgrid}{\tikz[remember picture, overlay]{%
    \foreach \x in {0, 600,..., 1800}
        \draw [line] (\dotposition{\x} + 3 pt, 9pt) -- (\dotposition{\x} + 3 pt, -3em) node[below, yshift=2pt, text=darkgray]{\tiny\x};
    \draw[red, dashed, line width=1] (\dotposition{\averageLoC} + 3 pt, 9pt) -- (\dotposition{\averageLoC} + 3 pt, -2.9em);}}
    
\newcommand{\LoCdots}[3]{\tikz[%
    dot/.style={circle,fill=#2,inner sep=1pt},
    triangle/.style={isosceles triangle, isosceles triangle apex angle=60, rotate=90, fill=#2,inner sep=0.9pt},
    diamond/.style={rectangle,rotate=45,fill=#2,inner sep=1.3pt}]{%
	\coordinate (origin) at (0,0);
	\coordinate (end) at (\LoCcolumnwidth pt, 0);
	\node at (origin) []{};
	\foreach \x in #1
	    \node at (\dotposition{\x} pt, \random)[#3]{};}}

\definecolor{very_lightgray}{RGB}{225, 225, 225}
\definecolor{simulation_col}{RGB}{29, 171, 61}
\definecolor{optimization_col}{RGB}{110, 55, 250}
\definecolor{ml_col}{RGB}{212, 102, 43}
\definecolor{codegreen}{rgb}{0,0.6,0}
\definecolor{codegray}{rgb}{0.5,0.5,0.5}
\definecolor{codepurple}{rgb}{0.58,0,0.82}
\definecolor{backcolour}{rgb}{0.95,0.95,0.92}
 
\lstdefinestyle{mystyle}{
    backgroundcolor=\color{backcolour},   
    commentstyle=\color{codegreen},
    keywordstyle=\color{magenta},
    numberstyle=\tiny\color{codegray},
    stringstyle=\color{codepurple},
    basicstyle=\footnotesize,
    breakatwhitespace=false,         
    breaklines=true,                 
    captionpos=b,                    
    keepspaces=true,                 
    numbers=left,                    
    numbersep=5pt,                  
    showspaces=false,                
    showstringspaces=false,
    showtabs=false,                  
    tabsize=2
}
\begin{document}




\title{Peel $\mid$ Pile? Cross-Framework Portability\newline  of Quantum Software\\
\thanks{MS, MF, and WM were supported by the German Federal Ministry of
Education and Research (BMBF), funding program ``quantum technologies---from
basic research to market'', grant number 13N15645. Our systematic search for open source quantum programming projects on GitHub using Google BigQuery, was supported by Google Cloud.}}

\author{\IEEEauthorblockN{Manuel Schönberger, Maja Franz}
\IEEEauthorblockA{
\textit{Technical University of}\\
\textit{Applied Sciences Regensburg,} Germany\\
\href{mailto:manuel.schoenberger@othr.de}{manuel.schoenberger@othr.de},\\
\href{mailto:maja.franz@st.othr.de}{maja.franz@st.othr.de}} 
\and
\and 
\IEEEauthorblockN{Stefanie Scherzinger}
\IEEEauthorblockA{\textit{Chair of Scalable Database Systems} \\
\textit{University of Passau}\\
Passau, Germany \\
\href{mailto:stefanie.scherzinger@uni-passau.de}{stefanie.scherzinger@uni-passau.de}}
\and
\IEEEauthorblockN{Wolfgang Mauerer}
\IEEEauthorblockA{\textit{Technical University of}\\
\textit{Applied Sciences Regensburg}\\
\textit{Siemens AG, Corporate Research}\\
\href{mailto:wolfgang.mauerer@othr.de}{wolfgang.mauerer@othr.de}}
}

\maketitle

\begin{abstract}
In recent years, various vendors have made quantum software frameworks available. 
Yet with vendor-specific frameworks, code portability seems at risk, especially in a field where 
hardware and software libraries have not yet reached a consolidated state, and even foundational
aspects of the technologies are still in flux.  Accordingly, the development of vendor-independent
quantum programming languages and frameworks is often suggested. 
This follows the established architectural pattern of introducing additional levels of abstraction
into software stacks, thereby \emph{piling on} layers of abstraction. Yet software architecture
also provides seemingly less abstract alternatives, namely to focus on hardware-specific formulations
of problems that \emph{peel off} unnecessary layers.
In this article, we quantitatively and experimentally explore these strategic alternatives, and
compare popular quantum frameworks from the software implementation perspective.
We find that for several specific, yet generalisable problems, the mathematical formulation
of the problem to be solved is not just sufficiently abstract and serves as precise description,
but is likewise concrete enough to allow for deriving framework-specific implementations with
little effort. Additionally, we argue, based on analysing dozens of existing quantum codes,
that porting between frameworks is actually low-effort, since the quantum- and framework-specific
portions are very manageable in terms of size, commonly in the order of mere hundreds of lines of code.
Given the \emph{current} state-of-the-art in quantum programming practice,
this leads us to argue in favour of \emph{peeling off} unnecessary abstraction levels. 

\end{abstract}

\section{Introduction}

In recent years, academia (\emph{e.g.},~\cite{Mauerer:2005}) and vendors have made frameworks for
developing quantum software available, such as IBM Qiskit~\cite{Qiskitdoc},
Pennylane~\cite{bergholm2020pennylane}, TensorFlow Quantum (TFQ)~\cite{Broughton.06.03.2020}, and
D-Wave Ocean~\cite{oceansdk}. Oftentimes, these vendors also provide access to quantum processing
units (QPUs). Since the current implementations of quantum algorithms are inherently coupled to the
framework used, the development of a vendor-independent quantum programming language as part of a
hardware-independent quantum processing framework has been suggested~\cite{FrankLeymann.2020}. A
high-level quantum programming language may moreover be beneficial from a quantum software
engineering perspective~\cite{zhao2020quantum,Krueger:2020}. Some quantum frameworks enabling
abstraction, to an
extent, are already available, like QC Ware's Quasar library~\cite{quasar} and the Atos Quantum
Learning Machine~\cite{atosqml}. Moreover, methods for automatically proposing suitable quantum
hardware for specific problems, formulated in a vendor-independent language, have been
proposed~\cite{FrankLeymann.2020, Weder.2021}. The idea of making quantum software development
hardware-independent and thus \emph{piling on} new layers of abstraction, is particularly attractive,
as both quantum hardware and software libraries are still evolving. More broadly, tools such as
GitHub Copilot~\cite{githubcopilot} successfully demonstrate how a higher degree of abstraction
increases developer efficiency: In Copilot, an AI engine proposes entire lines of code automatically.

Since quantum software development is still at an early stage, the research community still lacks
insights into the available systems and the specific properties that make them suitable for certain
quantum algorithms. Much like  QPUs themselves, these properties are evolving. Developing a
sufficiently accurate and future-proof automated selection process of quantum hardware is therefore
difficult to accomplish, at the current stage. 

Moreover, the actual benefits of further abstraction levels in quantum software development are still unclear: they strongly depend on the effort required for migrating existing implementations onto another framework. This effort needs to outweigh both the expense of creating an additional abstraction layer and the abstraction effort regarding existing implementations. Otherwise, the prospects for a vendor-independent programming language will be limited, as demonstrated by the lack of adoption of the programming language Ada, once developed with similar
aspirations~\cite{ichbiah1979rationale}.

We argue that we need to thoroughly understand the specific characteristics of the quantum frameworks. Previously, LaRose et al.~\cite{LaRose.2019} investigated four quantum frameworks and compared a variety of aspects, such as the available QPUs and library support. Moreover, a comparison framework for quantum frameworks was presented by Viez et al~\cite{Vietz.2021}. Yet we find the software implementation perspective underexplored.

We need to gain an understanding whether \emph{piling on} new layers of
abstraction is advisable, or whether we should rather focus our joint efforts on
hardware-specific implementations and thus \emph{peel off} unnecessary layers
instead. Accordingly, we systematically compare four popular quantum
frameworks~\cite{OwenLockwood.2020, Chen.2020,
Skolik.28.03.2021, franz22, Trummer.2016, Feld.2019, Khairy.2019}: (1) Qiskit, (2)
Pennylane, (3) TFQ, which incorporates Google's Cirq, and (4) D-Wave Ocean. 

\smallskip
\noindent
\textbf{Contributions.}
Our contributions are as follows: 
\begin{compactitem}
    \item We size up the quantum-specific code for several hybrid quantum-classical algorithms, all implemented by third parties. Our key insight is that the solutions studied require less than two thousand lines of code for encoding the quantum-specific parts. The quantum-specific code is thus small and manageable, comparable in volume to small personal programming projects. 
    \item  We compare the software development process given different development frameworks.
    We focus on two specific and rather novel application cases, namely reinforcement learning and  multi-query optimisation in databases engines. While reformulating these problems to quantum algorithms is conceptually challenging, we again find that the actual implementation effort for all frameworks is very manageable,
    to the point of straightforward.
    \item We compare the portability of quantum software across frameworks. Specifically,
    we assess how strongly an implementation is coupled to the underlying framework. While the
    framework-specific implementation effort varies between applications, only small-scale code
    portions are involved in general. This renders cross-framework porting a task that involves little effort.
\end{compactitem}

We then discuss the peel vs.\ pile trade-off given these results.

\smallskip
\noindent
\textbf{Structure.}
Chapter~\ref{promising_applications} introduces problem domains for which quantum computing is particularly promising, and investigates the code volume of the quantum-specific part of existing implementations. Chapter~\ref{sec:portability} studies cross-framework portability for two specific application use cases. Chapter~\ref{conclusion} concludes.

\section{Dimensions of Hybrid QC applications}

\label{promising_applications}

We size up the quantum-specific code for realistic, hybrid quantum-classical algorithms by considering applications
where quantum computing promises speedups.
The \emph{Quantum Application and Technology Consortium} (QuTAC) comprises ten multi-national companies from different sectors, in particular
automotive, chemistry, insurance, and technology.
The consortium identifies relevant
application domains~\cite{Bayerstadler.2021}: optimisation, machine learning (ML),
and simulation. 

\begin{table}[tb]
	\centering
	\caption{Quantum code volume of quantum-classical
	applications (files importing a 
	quantum framework library are considered quantum code). Dashed red line: Cross-domain average (\averageLoC).}\label{tab:relevance_appl_cases}
	\begin{tabular}{lp{\LoCcolumnwidth pt}l}
		\toprule
		\textbf{Domain} & \textbf{Lines of Code (LoC)} & \textbf{References}\\
		\midrule
		Optimisation & \LoCgrid\LoCdots{\optimizationLoC}{optimization_col}{triangle} 
		& \cite{Lucas.2014, Martonak.2004, Hogg.2003, Feld.2019, Gabor.2019, portfolioOpt, kruger2020quantum, sax2020approximate, antennaSel, maxCut} \\
		ML & \LoCdots{\mlLoC}{ml_col}{dot} 
		& \cite{OwenLockwood.2020, OwenLockwood.2020_GH, Skolik.28.03.2021, Skolik_GH, Chen.2020, Chen.2020_GH, mitarai2018quantum, hellstern2021analysis, farhi2018classification} \\
		Simulation & \LoCdots{\simulationLoC}{simulation_col}{diamond}
        & \cite{barison2020quantum, barison2020quantum_GH, copenhaver2021using_GH, copenhaver2021using} \\
		\bottomrule
	\end{tabular}
	
\end{table}

Table~\ref{tab:relevance_appl_cases} references previous work and code repositories related to these
problem domains. The material was collected by manually reviewing existing literature, but also by a
keyword search on a current snapshot of open source projects on GitHub, using Google BigQuery. These
and all further steps of our analysis are fully reproducible~\cite{Mauerer:2022} using our provided
reproduction package\footnote{Zenodo: \url{https://doi.org/10.5281/zenodo.5898296}}. The table lists
the total lines of code of all files within a project, where the files import some quantum framework
library. As such, this is a generous over-approximation of the share of the quantum-specific code.
Other possible metrics include code maintainability or readability. However, the former is hardly
applicable for software with only several hundred lines of code, whereas the latter is hard to
quantify. 

Similarly to classical software, the lines of code metric (LoC) has been suggested as useful for evaluating the size of quantum software and the process~\cite{ramsauer:2019} and development effort~\cite{Zhao.2021}. Other metrics correlate with lines of code~\cite{mamun.2017}, rendering them a suitable proxy metric. In the applications analysed, we also
found that the number of commits containing changes to quantum-related
code correlates with the LoC metric.

We observe that these numbers are small, which we attribute to the
representation pattern for many of these problems, particularly optimisation problems: typically, they are reformulated as quadratic unconstrained binary optimisation (QUBO) problems~\cite{Lucas.2014, Martonak.2004, Hogg.2003, Feld.2019, Gabor.2019}, to leverage existing implementations of quantum algorithms and solvers. Consequently, the bulk of development time is actually spent on finding suitable reformulations (rather than implementing complex control flow paths). This suggests that for these problems, the number of framework-specific and quantum-related implementation steps might be limited, since they mostly consist of calling existing quantum subroutines. Similar observations can be made for machine learning and simulation problems.

To verify or refute this indication, we analyse two specific and practically relevant problems representing two of the discussed domains in more detail. Specifically, we investigate implementation complexity and the quantum-specific steps.

\section{Cross-framework portability}
\label{sec:portability}

We critically evaluate and compare the gate-based frameworks Qiskit, Pennylane, TFQ, and the quantum annealing framework D-Wave Ocean, regarding the software implementation process. For our analysis, we choose two specific application scenarios that are subject to current research: (A)~reinforcement learning~\cite{Chen.2020, OwenLockwood.2020, Skolik.28.03.2021, franz22} and (B)~multi-query optimisation~\cite{Trummer.2016}. Due to the restriction to QUBO problems, we do not consider the Ocean framework for RL. More specifically, for each implementation we evaluate its size (in terms of lines of code) and its complexity with respect to the available documentation and library support.
Moreover, we investigate the framework-specific and quantum-related implementation steps, which determine how strongly the implementations are tied to the quantum frameworks. Based on our findings, we discuss the benefits a new abstraction layer may provide. 


%

\def\picwidth{18cm}
\def\picheight{9cm}
\def\innerwidth{4.0cm}
\def\hsep{0.2cm}
\def\vsep{0.1cm}
\newlength\boxwidth\setlength{\boxwidth}{(\picwidth-\innerwidth)/2-\hsep}
\newlength\boxheight\setlength{\boxheight}{(\picheight-\vsep-\vsep)/3}
\algnewcommand{\LeftComment}[1]{\State \(\triangleright\) #1}
\definecolor{qiskitcol}{RGB}{112,48,160}
\definecolor{pennylanecol}{RGB}{112,173,71}
\definecolor{tfqcol}{RGB}{255,192,0}

\newcommand{\pwidth}{0.24\linewidth}
\newcommand{\pchcorr}{\hspace*{-0.9em}}
\newcommand{\algsize}{\fontsize{6.6}{7.6}\selectfont}
\newcommand{\gendeepq}{\pchcorr\begin{minipage}{\pwidth}
\renewcommand*\ttdefault{cmvtt} 
\algsize\begin{algorithmic}
    \LeftComment{\textbf{Quantum Deep Q-Learning}}
    \State Init replay buf $\mathbb{D}$, (target) VQC $\theta^{(-)}$
\For{$s\gets 0, \texttt{steps}$}
    \State Sample $\mathbb{B}=$
    \State \hspace*{2em}$(s_t, a_t, s_{t+1}, r_{t+1})\backslash\mathbb{D}$
    \ForAll{$b_{i} \in \mathbb{B}$}
        \LeftComment{Target}
        \State $y_i \gets \gamma \max\limits_{a'} Q(s_{i+1}, a'; \theta^-)$ 
        \State $q_i \gets Q(s_i, a_i; \theta)$ \Comment{Q-Value}
    \EndFor 
    \State $L(\theta) \gets (y - q)^2$ \Comment{Loss}
    \State Update $\theta$ (param shift rule~\cite{mitarai2018quantum, schuld2019evaluating})
    \If{$s\bmod{\texttt{update}} = 0$}
    \State $\theta^{-} \gets \theta$
    \EndIf 
\EndFor 
\end{algorithmic}
\end{minipage}}


\newcommand{\genqaoa}{\pchcorr\begin{minipage}{\pwidth}
\renewcommand*\ttdefault{cmvtt}
\algsize\begin{algorithmic}
    \LeftComment{\textbf{Multi Query Optimization}}
    \State Init MQO QUBO $qm$, circ.\ params $\beta$, $\gamma$
\State $c_{\texttt{lin}} \texttt{,} c_{\texttt{qdr}} \gets \texttt{IsingCoeffs}(qm)$
\State $hl \gets \texttt{\{\}}$
\For{$k \gets 0, p$}
    \State $hl \gets hl\frown\texttt{CostHam(}c_{\texttt{lin}}, c_{\texttt{qdr}}\texttt{)}$
    \State $hl \gets hl\frown\texttt{MixerHam()}$
\EndFor 
\State $qc \gets \texttt{buildQCircuit(}hl, \beta, \gamma$ \texttt{)}
\State Initialize classical optimizer $opt$
\While{$\lnot$converged}
    \State $\beta, \gamma \gets opt\texttt{.step(}qc, \beta, \gamma \texttt{)}$ 
\EndWhile 
\State $r \gets \texttt{sample(}qc, \beta, \gamma \texttt{)}$
\end{algorithmic}\end{minipage}}

\newcommand{\codefont}{\fontsize{3.5}{4}\selectfont}
\lstset{language=python,basicstyle=\codefont\ttfamily,numbers=none, 
  backgroundcolor={},framesep=0pt,framerule=0pt,xleftmargin=0pt,tabsize=1,
  showtabs=true,showspaces=false}

\newcommand{\lsttopcorr}{\vspace*{-0.75em}}
\newsavebox\qiskitla\newsavebox\qiskitlb
\begin{lrbox}{\qiskitla}
\begin{minipage}{(\boxwidth-\hsep)/2}\lsttopcorr
\begin{lstlisting}
# define circuit
class VQC_Layer(Module):
  def __init__(self, n_qubits, n_layers, shots, device):
    self.circuit = QuantumCircuit(n_qubits)
    # input part
    for i, input in enumerate(input_params):
      self.circuit.rx(input, i)
    for i in range(n_layers):
      self.generate_layer(weight_params[i*n_qubits*2 : (i+1)*n_qubits*2])
    readout_op = ListOp([
      ~StateFn(pauli_op_list([('ZZII', 1.0)])) @ StateFn(self.circuit),
      ~StateFn(pauli_op_list([('IIZZ', 1.0)])) @ StateFn(self.circuit)])
    qnn = OpflowQNN(readout_op,
       input_params=input_params,
       weight_params=weight_params,
\end{lstlisting}\vspace*{-1em}\end{minipage}
\end{lrbox}
\begin{lrbox}{\qiskitlb}
\begin{minipage}{(\boxwidth-\hsep)/2}\lsttopcorr
\begin{lstlisting}
       quantum_instance=qi,
       gradient=Gradient()
    self.qnn = TorchConnector(qnn, initial_weights=torch.Tensor(np.zeros(n_qubits*n_layers*2)))
    
  def generate_layer(self, params):
    # variational part
    for i in range(self.n_qubits):
        self.circuit.ry(params[i*2], i)
        self.circuit.rz(params[i*2+1], i)
    # entangling part
    for i in range(self.n_qubits):
        self.circuit.cz(i, (i+1) % self.n_qubits)
 
    # Q-value calculation
  def forward(self, inputs):
    return self.qnn(inputs)
\end{lstlisting}\vspace*{-1em}\end{minipage}
\end{lrbox}

\newsavebox\pennyla\newsavebox\pennylb
\begin{lrbox}{\pennyla}
\begin{minipage}{(\boxwidth-\hsep)/2}\lsttopcorr
\begin{lstlisting}
# define circuit and Q-value calculation
@qml.qnode(device, wires=config.n_qubits), interface='tf', diff_method='parameter-shift')
def circuit(inputs, weights):
  # input part
  for i in range(config.n_qubits):
    qml.RX(inputs[i], wires=i)

  for i in range(config.n_layers):
    generate_layer(weights[i], config.n_qubits)
  return [qml.expval(qml.PauliZ(0) @ qml.PauliZ(1)), qml.expval(qml.PauliZ(2) @ qml.PauliZ(3))]
\end{lstlisting}\vspace*{-1em}\end{minipage}
\end{lrbox}
\begin{lrbox}{\pennylb}
\begin{minipage}{(\boxwidth-\hsep)/2}\lsttopcorr
\begin{lstlisting}
def generate_layer(params, n_qubits): 
  # variational part
  for i in range(n_qubits):
    qml.RY(params[i][0], wires=i)
    qml.RZ(params[i][1], wires=i)
  # entangling part
  for i in range(n_qubits):
    qml.CZ(wires=[i, (i+1) % n_qubits])

# initialize Keras Layer
VQC_Layer = qml.qnn.KerasLayer(
  qnode=circuit,
  weight_shapes={'weights': (config.n_layers, config.n_qubits, 2)},
  weight_specs = {"weights": {"initializer": "Zeros"}},
  output_dim=2,
  name='VQC_Layer')
\end{lstlisting}\vspace*{-1em}\end{minipage}
\end{lrbox}

\newsavebox\tqla\newsavebox\tqlb
\begin{lrbox}{\tqla}
\begin{minipage}{(\boxwidth-\hsep)/2}\lsttopcorr
\begin{lstlisting}
# define circuit
class VQC_Layer(keras.layers.Layer):
def __init__(self, n_qubits, n_layers):
  # ... definition of variables 
  circuit = cirq.Circuit()
  circuit.append([cirq.rx(inputs[i]).on(qubit) for i, qubit in enumerate(self.qubits)])
  for i in range(n_layers):
    circuit.append(
      self.generate_layer(params[i*n_qubits*2 : (i+1)*n_qubits*2]))
  readout_op = [
    cirq.PauliString(cirq.Z(qubit) for qubit in self.qubits[:2]),
    cirq.PauliString(cirq.Z(qubit) for qubit in self.qubits[2:])]
  self.vqc = tfq.layers.ControlledPQC(circuit, readout_op, 
    differentiator=ParameterShift())
\end{lstlisting}\vspace*{-1em}\end{minipage}
\end{lrbox}
\begin{lrbox}{\tqlb}
\begin{minipage}{(\boxwidth-\hsep)/2}\lsttopcorr
\begin{lstlisting}
def generate_layer(self, params):
  circuit = cirq.Circuit()
  # variational part
  for i, qubit in enumerate(self.qubits):
    circuit.append([
      cirq.ry(params[i*2]).on(qubit),
      cirq.rz(params[i*2+1]).on(qubit)])
  # entangling part
  for i in range(self.n_qubits):
    circuit.append(cirq.CZ.on(self.qubits[i], 
      self.qubits[(i+1) % self.n_qubits]))
  return circuit

# Q-value calculation
def call(self, inputs):
  # ... classical input processing
  return self.vqc([tiled_up_circuits, joined_vars])
\end{lstlisting}\vspace*{-1em}\end{minipage}
\end{lrbox}

\newsavebox\qiskitra\newsavebox\qiskitrb
\begin{lrbox}{\qiskitra}
\begin{minipage}{(\boxwidth-\hsep)/2}\lsttopcorr
\begin{lstlisting}
def construct_model(model,queries,
                    costs,savings):
  v = model.binary_var_list(len(costs))
  epsilon = 0.25
  wl= calculate_wl(costs, epsilon)
  wm= calculate_wm(savings, wl)
  El=model.sum(-1*(wl-costs[i])*v[i] \
     for i in range(0, len(costs)))
  Em=model.sum(model.sum(wm*v[i]*v[j] \
     for (i,j) in itertools.combinations
                  (queries[k], 2)) \
     for k in queries.keys())
  Es=model.sum(-s*v[i]*v[j] \
     for ((i,j), s) in savings)
  return(El + Em + Es)
\end{lstlisting}\vspace*{-1em}\end{minipage}
\end{lrbox}
\begin{lrbox}{\qiskitrb}
\begin{minipage}{(\boxwidth-\hsep)/2}\lsttopcorr
\begin{lstlisting}
def solve_with_QAOA(qubo):
  qmeas=qiskit.algorithms.QAOA(
    quantum_instance=Aer.get_backend('qasm_simulator'), 
    initial_point=[0., 0.])
  qaoa=MinimumEigenOptimizer(qaoa_meas)
  qres=qaoa.solve(qubo)
  return qres,qmeas.get_optimal_circuit()
	
def solve_MQO(queries, costs, savings):    
  model=Model('docplex_model')
  model.minimize(construct_model(
      model, queries, costs, savings))
  qubo=QuadraticProgram()
  qubo.from_docplex(model)
  result_QAOA, QAOA_circuit=solve_with_QAOA(qubo)
\end{lstlisting}\vspace*{-1em}\end{minipage}
\end{lrbox}

\newsavebox\pennyra\newsavebox\pennyrb
\begin{lrbox}{\pennyra}
\begin{minipage}{(\boxwidth-\hsep)/2}\lsttopcorr
\begin{lstlisting}
def get_ising_model(queries, costs,
                    savings):
  model = Model('docplex_model')
  v = model.binary_var_list(len(costs))
  epsilon = 0.25
  wl = calculate_wl(costs, epsilon)
  wm = calculate_wm(savings, wl)
  El = model.sum(-1*(wl-costs[i])*v[i] \
       for i in range(0, len(costs)))
  Em = model.sum(model.sum(wm*v[i]*v[j] \
       for (i,j) in itertools.combinations(
          queries[k], 2)) for k in queries.keys())
  Es = model.sum(-s*v[i]*v[j] \
       for ((i,j), s) in savings)
  model.minimize(El+Em+Es)  
  qubo = translators.from_docplex_mp(model)
  return qubo.to_ising()	
\end{lstlisting}\vspace*{-1em}\end{minipage}
\end{lrbox}
\begin{lrbox}{\pennyrb}
\begin{minipage}{(\boxwidth-\hsep)/2}\lsttopcorr
\begin{lstlisting}
def solve_MQO(queries, costs, savings, p=1):   
  sing, offset = get_ising_model(
                    queries, costs, savings)
  wires = range(len(costs))
  cham = create_cost_hamiltonian(linear, quadratic, offset)
  mham = create_mixer_hamiltonian(wires)
  dev = qml.device("default.qubit", wires=wires)
  circuit = qml.QNode(cost_function, dev)   
  params = get_initial_params(p)  
  optimizer = qml.GradientDescentOptimizer()
  steps = 200  
  for i in range(steps):
    params = optimizer.step(circuit, params,
                wires=wires, depth=p,
                cost_hamiltonian=cham,
                mixer_hamiltonian=mham)
\end{lstlisting}\vspace*{-1em}\end{minipage}
\end{lrbox}

\newsavebox\tqra\newsavebox\tqrb
\begin{lrbox}{\tqra}
\begin{minipage}{(\boxwidth-\hsep)/2}\lsttopcorr
\begin{lstlisting}
def get_ising_model(queries,costs,savings):
  model = Model('docplex_model')
  v = model.binary_var_list(len(costs))
  epsilon = 0.25
  wl = calculate_wl(costs, epsilon)
  wm = calculate_wm(savings, wl)
  El = model.sum(-1*(wl-costs[i])*v[i] for i in range(0, len(costs)))
  Em = model.sum(model.sum(wm*v[i]*v[j] \
            for (i,j) in itertools.combinations(queries[k], 2)) 
            for k in queries.keys())
  Es = model.sum(-s*v[i]*v[j] for ((i,j), s) in savings)
  model.minimize(El+Em+Es)  
  qubo = translators.from_docplex_mp(model)
  return qubo.to_ising()
def solve_MQO(queries, costs, savings, p=1):
  ising, offset = get_ising_model(queries, costs, savings)
\end{lstlisting}\vspace*{-1em}\end{minipage}
\end{lrbox}
\begin{lrbox}{\tqrb}
\begin{minipage}{(\boxwidth-\hsep)/2}\lsttopcorr
\begin{lstlisting}
  coeffs = np.real(ising.primitive.coeffs)
  pauli_array = ising.primitive.settings['data'].array
  linear, quadratic = get_coefficients_from_Pauli_array(pauli_array, coeffs)
  cirq_qubits = cirq.GridQubit.rect(1, len(costs))
  # ... Init parameters and hamiltonians ...
  qaoa_circuit = tfq.util.exponential(operators=hamiltonians,
                          coefficients=qaoa_parameters)
  hadamard_circuit = get_hadamard_circuit(cirq_qubits)
  # ... Initialize Keras Model ...
  model.add(tfq.layers.PQC(model_circuit,
            model_readout, backend=cirq.Simulator()))
  # ... Train the model ...
\end{lstlisting}\vspace*{-1em}\end{minipage}
\end{lrbox}

\begin{figure*}
    \input{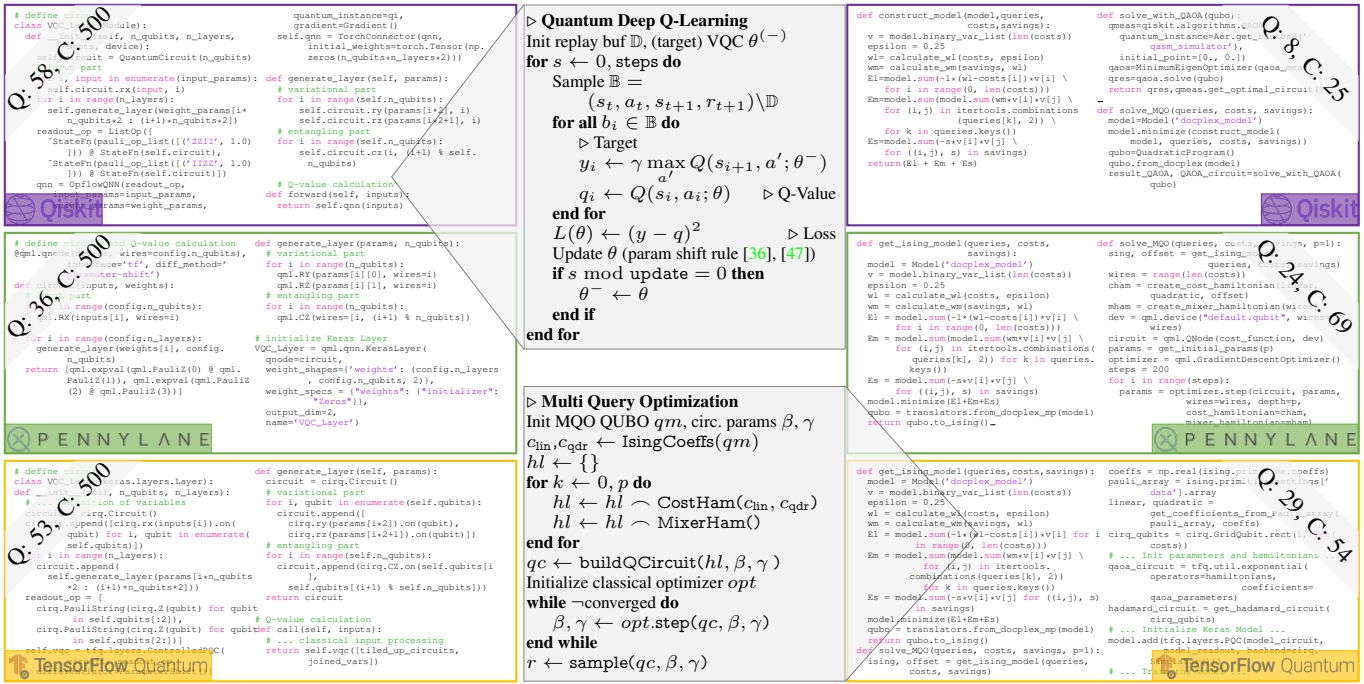}\vspace*{-2em}
    \caption{(Best viewed in colour.) Pseudocode capturing the essential structure
    of quantum algorithms for reinforcement learning (top centre) and multi-query optimisation (bottom centre), and their concrete implementations in three frameworks (left: RL, right: MQO), together with approximate lines of code for quantum-specific components (Q:~\(\langle N\rangle\)), and classical (C:~\(\langle N\rangle\)) contributions (based on
    a manual classification by the authors). The python code, although functional, is not meant to be read, but merely gives a sense of scale.
    Regardless
    of the framework, the concrete implementations are close in size to the abstract pseudo-code representation, indicating that further abstraction
    layers or domain-specific quantum programming languages have very limited potential for additional reduction in size, and increase in expressivity.
    We only show classical code that is directly interrelated with quantum code components in the figure.}
    \label{fig:framework_comparison}
\end{figure*}

\subsection{Reinforcement Learning}
Most reinforcement learning (RL) formulations centre around a Markov Decision Process (MDP)~\cite{Bellman1957}:
An \emph{agent} interacts with an \textit{environment} to maximise a cumulative reward $G_t = \sum_{t'=t}^T \gamma^{t'} R_{t'}$
until a terminal state $S_T$ is reached, with $R_{t'}$ being the reward at time step~$t'$ and a discount
factor~$\gamma$~\cite{sutton2018}. We focus on Deep Q-Learning~\cite{WatkinsDayan92, mnih2013playing}, where
the idea is to learn the optimal \textit{action-value function}, also referred to as \textit{Q-function}:
$Q_*(s,a) = \max_\pi \mathbb{E} \left[G_t | S_t = s, A_t = a, \pi \right]$. It represents the
return, or accumulative reward~$G_t$, expected when taking an action $a$ in the environment's state~$s$,
then following a policy $\pi$ in future states. An optimal policy $\pi_*$ can be recovered by taking
the action that maximises future \textit{Q-values}: $\pi_*(s) = \arg \max_{a} Q_*(s,a)$. In classical
Deep Q-Learning, this is achieved by training a neural network to satisfy the 
\textit{Bellman Optimality Equation}~\cite{Bellman1957} that relates the values of a state-action pair to the value of the next state:
\begin{align}
    Q_*(s,a) = \mathbb{E}\Big[R_t + \gamma \max_{a'} Q_*(S_{t+1}, a') \mid &S_t = s, \nonumber &A_t = a\Big]
\end{align}

In quantum-based RL, the neural network can be replaced by a \emph{variational quantum circuit}
(VQC)~\cite{mitarai2018quantum, Chen.2020, Skolik.28.03.2021, OwenLockwood.2020}, parameterised by weights $\theta$.
The algorithm sketched in Figure~\ref{fig:framework_comparison} (top centre) employs the Double Q-Learning
approach as suggested in~\cite{van2016deep}, which calculates targets with a \textit{target network}
or a \textit{target VQC} in the quantum domain, parameterised by~$\theta^-$.


As a first step, we implemented classical RL with a neural network, using TensorFlow~\cite{tensorflow_developers_2021_5645375} and PyTorch~\cite{pytorch}. We chose these frameworks because Pennylane offers an interface for both, TFQ builds upon TensorFlow,
and Qiskit provides a machine learning library based on PyTorch. Since common quantum Q-Learning
algorithms~\cite{Chen.2020, Skolik.28.03.2021, OwenLockwood.2020} do not examine annealing approaches, we only consider
gate-based frameworks.

After confirming the correctness of our implementations, we replaced the classical neural network
with a VQC based on standard framework patterns, in particular using python to represent
a VQC for machine learning with trainable parameters. It essentially comprises three 
code elements: (1)~the VQC definition, (2)~calculation and processing of Q-values, 
and (3)~calculation of gradients in a quantum-classical back-propagation
procedure~\cite{mitarai2018quantum, schuld2019evaluating}. All frameworks offer similar library
classes for this purpose. In each case, we could easily isolate the quantum-based steps from the classical
algorithm.

As we discuss in Sec.~\ref{sec:lineup}, the framework-specific implementation 
steps are almost interchangeable\footnote{We did observe major differences in run-time: Calculating
gradients for one batch takes \numprint{47748} ms in Qiskit, \numprint{1212} ms in Pennylane,
and \numprint{659} ms in TFQ; the differences substantially impact practical utility (measurements
are averaged over 100 batches and conducted on the same device.)}.
An abstraction layer at this level could be beneficial, since TFQ is coupled with TensorFlow,
and Qiskit with PyTorch. Hence, it could ease porting between ML and quantum frameworks.

\subsection{Multi Query Optimisation}
Multi query optimisation (MQO) is a longstanding problem in database research.
It seeks to determine a globally optimal set of execution plans for a batch of database queries, minimising the
overall execution cost by reusing common subexpressions~\cite{DBLP:reference/db/Roy018}. The problem has been addressed on
a D-Wave quantum annealer, based on a reformulation into a QUBO problem~\cite{Trummer.2016}. QUBO problems and their
equivalent Ising formulations~\cite{Bian.2010} can also be solved on gate-based QPUs with variational
hybrid quantum-classical algorithms~\cite{McClean.2016}, such as the quantum approximate optimisation
algorithm (QAOA)~\cite{Farhi.2014}. We can solve MQO problems on gate-based
frameworks~\cite{schoenberger2021quantumqueryopt, fankhauser2021multiple} and therefore on all considered quantum frameworks
using the reformulation approach presented in Ref.~\cite{Trummer.2016}.

We again discuss the implementation procedure for each framework. The Ocean implementation for D-Wave 
is straightforward---it suffices to apply the QUBO reformulation proposed in~\cite{Trummer.2016}. Using
framework-provided classes, we create a quadratic model that serves as input for all solvers.
The Ocean implementation is the most compact among all frameworks.


For gate-based frameworks, we use the same reformulation approach to solve the problem with QAOA. The algorithm is sketched in Figure~\ref{fig:framework_comparison} (bottom centre). Here, we first searched the available documentation and libraries for artefacts related to QAOA. We found that Qiskit offers a library that fully encapsulates all QAOA steps. Much like the Ocean implementation, Qiskit requires only a limited number of steps: We use the IBM DOcplex tool~\cite{docplex} to apply the QUBO transformation using mathematical expressions. We then use classes and methods provided by Qiskit to create a quadratic model based on the DOcplex model. Finally, we determine an optimal solution for the quadratic problem using a provided optimiser, which transforms the QUBO to an Ising model and which we configure to internally use the available QAOA solver. No explicit QAOA circuit specification is necessary for Qiskit.

We did not find any comparable libraries for Pennylane and TFQ. However, for Pennylane, a library containing utility functions for QAOA (e.g., for applying the cost and mixer layers) exists, which simplifies the process of creating the QAOA circuit. In both cases, we create QAOA circuits by simply following demo code from the documentation. 

The QAOA circuit consists of an alternating sequence of repeating cost and mixing operators, where the number of repetitions is given by a parameter $p$~\cite{Farhi.2014}. To create the cost Hamiltonian needed for the cost operators, we need the Ising coefficients of our problem formulation. Therefore, we base the Pennylane and TFQ implementations on our Qiskit implementation, which allows us to use available methods for converting QUBOs into an Ising model. As before, the framework-specific steps are few and simple: they consist of creating the operators and parameterised quantum circuits based on the Ising coefficients. The parameters are classically optimised in an outer loop (e.g., by gradient descent methods).


We found weak framework coupling for all MQO implementations. Ocean and Qiskit
mostly require QUBO transformation, and resulting models serve as input for the solvers offered by frameworks.
The implementations for Qiskit, Pennylane, and TFQ are near-identical up to determining the QUBO model
or the respective Ising coefficients. The remaining framework-specific steps for Pennylane and TFQ are largely independent
of the concrete problem, and moreover straightforward to implement through the use of existing libraries and available
documentation. The effort for porting the MQO implementation across frameworks
is minor, and an additional abstraction layer at implementation level provides little benefit.

\subsection{Application Scenario Lineup}
\label{sec:lineup}

Figure~\ref{fig:framework_comparison} shows the quantum-relevant code for the 
use-cases in the gate-based frameworks. It allows for side-by-side comparison.
In the centre, we show the abstract pseudo-code for RL and MQO.
To the left and right, we show quantum-specific code for each framework
(the python code is not meant to be readable, but merely to provide a sense of scale).

The implementations are of limited LoC size,
and well below multi-million LoC typically considered in software engineering~\cite{Bass:2012}.
In fact, the quantum-specific python code ranges between just 8 and 58 lines of code.

Most importantly, they are comparable in size to the mathematical pseudo-code
representation. This suggests limited potential for further abstraction. Porting between
frameworks is mostly a direct substitution of APIs without
structural code changes, indicating that the expressivity is essentially optimal.

\section{Discussion and conclusion}
\label{conclusion}
Programming quantum computers is, at the current state of technology, often perceived as a very low-level
task, comparable to programming early-generation classical machines. We have studied potentials and limitations
for extending the state-of-the-art with higher-level abstractions and device-independent presentation
of quantum algorithms using two means: (a)~by learning from existing quantum programs, and (b)~by implementing
two advanced use-cases for multiple, popular quantum programming frameworks, and judging similarities
across frameworks, with an abstract pseudo-code representation.
For RL, we isolated the quantum-specific implementation details from the classical algorithm. 
For MQO, we used a QUBO reformulation applicable for all frameworks.

In all cases, quantum-specific portions are small, and the level of abstraction is not much different
between pseudo-code and all frameworks. We see no reason to assume much difference between the considered problems and others of similar size and problem domains, in this regard. All scenarios are orders of magnitude away from problem sizes 
considered challenging in software architecture and engineering practice and research.


Our findings suggest that in general, introducing new abstraction layers by crafting framework-independent
programming languages holds limited promise.
\nop{
We have conducted a systematic comparison of the popular quantum frameworks Qiskit, Pennylane, TFQ and D-Wave Ocean
from a software implementation perspective. Specifically, we have compared implementations for two application cases:
Reinforcement learning and multi-query optimisation. 

For RL, we isolated the quantum-specific implementation details from the classical algorithm. Given machine learning framework support, we could easily interchange quantum framework-specific steps, using the available documentation. Since TFQ and Qiskit are coupled to different machine learning frameworks, porting across these quantum frameworks  requires porting across machine learning frameworks. Therefore, introducing an additional abstraction layer between machine learning and quantum frameworks could indeed close a gap.

For MQO, we applied a reformulation into a QUBO problem, which allows us to solve such problems on all considered frameworks. We found that in all cases, the framework-specific implementation steps are largely independent of the considered problem, leading to a weak framework coupling. Using existing documentation and in some cases available libraries, the code can easily be ported. As such, we find that the mathematical QUBO formulation serves as a sufficient abstraction layer, rendering new abstraction layers unnecessary.

Our results show that the benefits of developing a vendor-independent quantum programming language and piling on a new abstraction layer highly depend on the application use case. 
Future research is required, to study further application cases regarding  portability.
} 
Still, our selected application cases represent problem domains which are considered promising candidates for
quantum speedups. Other problems of these domains are usually solved with similar patterns and paradigms. For instance,
optimisation problems are typically reformulated to leverage established quantum algorithms. This might change once
new quantum algorithms and paradigms are discovered. However, progress related to quantum algorithms has been
moderate---for instance, key algorithms like Grover search~\cite{Grover.1996} have been known for more than two
decades. Therefore, the familiar quantum patterns and paradigms are likely to persist for the foreseeable future. 

Ultimately, when deciding between piling on new abstraction layers or peeling off existing ones,
our results suggest the latter.

\nop{
\todo[inline]{    
     kleine studie zeigt, Code im kleinen Umfang,
     abstraction als pattern bietet sich nicht an,
     additional abstraction layer provides little benefits
     unless radically new approaches in q alg construction
     emerge
    
     on algorithm side, few new contributions in last few decades
     (popular Grover already celebrates 25th anniversary)
    
     tradeoff pile on/peel off, indicators are all towards the latter
     }
} 

\bibliographystyle{IEEEtran}
\bibliography{literature}

\begin{thebibliography}{10}
\providecommand{\url}[1]{#1}
\csname url@samestyle\endcsname
\providecommand{\newblock}{\relax}
\providecommand{\bibinfo}[2]{#2}
\providecommand{\BIBentrySTDinterwordspacing}{\spaceskip=0pt\relax}
\providecommand{\BIBentryALTinterwordstretchfactor}{4}
\providecommand{\BIBentryALTinterwordspacing}{\spaceskip=\fontdimen2\font plus
\BIBentryALTinterwordstretchfactor\fontdimen3\font minus
  \fontdimen4\font\relax}
\providecommand{\BIBforeignlanguage}[2]{{%
\expandafter\ifx\csname l@#1\endcsname\relax
\typeout{** WARNING: IEEEtran.bst: No hyphenation pattern has been}%
\typeout{** loaded for the language `#1'. Using the pattern for}%
\typeout{** the default language instead.}%
\else
\language=\csname l@#1\endcsname
\fi
#2}}
\providecommand{\BIBdecl}{\relax}
\BIBdecl

\bibitem{Mauerer:2005}
\BIBentryALTinterwordspacing
W.~Mauerer, ``Semantics and simulation of communication in quantum
  programming,'' 2005. [Online]. Available:
  \url{https://arxiv.org/abs/quant-ph/0511145}
\BIBentrySTDinterwordspacing

\bibitem{Qiskitdoc}
\BIBentryALTinterwordspacing
{IBM}, ``Qiskit: An open-source framework for quantum computing,'' 2021.
  [Online]. Available: \url{https://qiskit.org/}
\BIBentrySTDinterwordspacing

\bibitem{bergholm2020pennylane}
\BIBentryALTinterwordspacing
V.~Bergholm, J.~Izaac, M.~Schuld, C.~Gogolin, M.~S. Alam, S.~Ahmed, J.~M.
  Arrazola, C.~Blank, A.~Delgado, S.~Jahangiri, K.~McKiernan, J.~J. Meyer,
  Z.~Niu, A.~Száva, and N.~Killoran, ``Pennylane: Automatic differentiation of
  hybrid quantum-classical computations,'' 2020. [Online]. Available:
  \url{https://arxiv.org/abs/1811.04968}
\BIBentrySTDinterwordspacing

\bibitem{Broughton.06.03.2020}
\BIBentryALTinterwordspacing
M.~Broughton, G.~Verdon, T.~McCourt, A.~J. Martinez, J.~H. Yoo, S.~V. Isakov,
  P.~Massey, R.~Halavati, M.~Y. Niu, A.~Zlokapa, E.~Peters, O.~Lockwood,
  A.~Skolik, S.~Jerbi, V.~Dunjko, M.~Leib, M.~Streif, D.~von Dollen, H.~Chen,
  S.~Cao, R.~Wiersema, H.-Y. Huang, J.~R. McClean, R.~Babbush, S.~Boixo,
  D.~Bacon, A.~K. Ho, H.~Neven, and M.~Mohseni, ``Tensorflow quantum: A
  software framework for quantum machine learning.'' [Online]. Available:
  \url{https://arxiv.org/abs/2003.02989}
\BIBentrySTDinterwordspacing

\bibitem{oceansdk}
\BIBentryALTinterwordspacing
{D-Wave Systems Inc}, ``{Documentation for the Ocean SDK for solving problems
  on D-Wave quantum computers},'' 2021. [Online]. Available:
  \url{https://docs.ocean.dwavesys.com/en/stable/}
\BIBentrySTDinterwordspacing

\bibitem{FrankLeymann.2020}
{Frank Leymann}, {Johanna Barzen}, {Michael Falkenthal}, {Daniel Vietz}, and
  {Karoline Wild}, ``Quantum in the cloud: Application potentials and research
  opportunities,'' in \emph{Proceedings of the 10th International Conference on
  Cloud Computing and Services Science}, 2020, pp. 9--24.

\bibitem{zhao2020quantum}
\BIBentryALTinterwordspacing
J.~Zhao, ``Quantum software engineering: Landscapes and horizons,'' 2020.
  [Online]. Available: \url{https://arxiv.org/abs/2007.07047}
\BIBentrySTDinterwordspacing

\bibitem{Krueger:2020}
\BIBentryALTinterwordspacing
T.~Kr\"{u}ger and W.~Mauerer, \emph{Quantum Annealing-Based Software
  Components: An Experimental Case Study with SAT Solving}.\hskip 1em plus
  0.5em minus 0.4em\relax New York, NY, USA: Association for Computing
  Machinery, 2020, p. 445–450. [Online]. Available:
  \url{https://doi.org/10.1145/3387940.3391472}
\BIBentrySTDinterwordspacing

\bibitem{quasar}
\BIBentryALTinterwordspacing
{QC Ware}, ``Quasar library for creating and evaluating quantum computing
  circuits,'' 2022. [Online]. Available:
  \url{https://qcware-quasar.readthedocs.io/en/latest/}
\BIBentrySTDinterwordspacing

\bibitem{atosqml}
\BIBentryALTinterwordspacing
{Atos}, ``Quantum learning machine,'' 2020. [Online]. Available:
  \url{https://atos.net/wp-content/uploads/2020/11/Quantum-Learning-Machine-Brochure.pdf}
\BIBentrySTDinterwordspacing

\bibitem{Weder.2021}
B.~Weder, J.~Barzen, F.~Leymann, and M.~Salm, ``Automated quantum hardware
  selection for quantum workflows,'' \emph{Electronics}, vol.~10, no.~8, p.
  984, 2021.

\bibitem{githubcopilot}
\BIBentryALTinterwordspacing
A.~Ziegler, ``{A first look at rote learning in GitHub Copilot suggestions},''
  2021. [Online]. Available:
  \url{https://docs.github.com/en/github/copilot/research-recitation}
\BIBentrySTDinterwordspacing

\bibitem{ichbiah1979rationale}
J.~D. Ichbiah, B.~Krieg-Brueckner, B.~A. Wichmann, J.~G. Barnes, O.~Roubine,
  and J.-C. Heliard, ``Rationale for the design of the {Ada} programming
  language,'' \emph{ACM Sigplan notices}, vol.~14, no.~6b, pp. 1--261, 1979.

\bibitem{LaRose.2019}
R.~LaRose, ``Overview and comparison of gate level quantum software
  platforms,'' \emph{Quantum}, vol.~3, p. 130, 2019.

\bibitem{Vietz.2021}
D.~Vietz, J.~Barzen, F.~Leymann, and K.~Wild, ``On decision support for quantum
  application developers: Categorization, comparison, and analysis of existing
  technologies,'' in \emph{Proceedings ICCS 2021}, 2021, pp. 127--141.

\bibitem{OwenLockwood.2020}
{Owen Lockwood} and {Mei Si}, ``Reinforcement learning with quantum variational
  circuit,'' \emph{Proceedings of the AAAI Conference on Artificial
  Intelligence and Interactive Digital Entertainment}, vol.~16, no.~1, pp.
  245--251, 2020.

\bibitem{Chen.2020}
S.~Y.-C. Chen, C.-H.~H. Yang, J.~Qi, P.-Y. Chen, X.~Ma, and H.-S. Goan,
  ``Variational quantum circuits for deep reinforcement learning,'' \emph{IEEE
  Access}, vol.~8, pp. 141\,007--141\,024, 2020.

\bibitem{Skolik.28.03.2021}
\BIBentryALTinterwordspacing
A.~Skolik, S.~Jerbi, and V.~Dunjko, ``Quantum agents in the gym: a variational
  quantum algorithm for deep q-learning.'' [Online]. Available:
  \url{https://arxiv.org/abs/2103.15084}
\BIBentrySTDinterwordspacing

\bibitem{franz22}
\BIBentryALTinterwordspacing
M.~Franz, L.~Wolf, M.~Periyasamy, C.~Ufrecht, D.~D. Scherer, A.~Plinge,
  C.~Mutschler, and W.~Mauerer, ``Uncovering instabilities in
  variational-quantum deep q-networks,'' 2022. [Online]. Available:
  \url{https://arxiv.org/abs/2202.05195}
\BIBentrySTDinterwordspacing

\bibitem{Trummer.2016}
I.~Trummer and C.~Koch, ``Multiple query optimization on the {D-Wave 2X}
  adiabatic quantum computer,'' \emph{Proceedings of the VLDB Endowment},
  vol.~9, no.~9, pp. 648--659, 2016.

\bibitem{Feld.2019}
S.~Feld, C.~Roch, T.~Gabor, C.~Seidel, F.~Neukart, I.~Galter, W.~Mauerer, and
  C.~Linnhoff-Popien, ``A hybrid solution method for the capacitated vehicle
  routing problem using a quantum annealer,'' \emph{Frontiers in ICT}, vol.~6,
  p.~13, 2019.

\bibitem{Khairy.2019}
\BIBentryALTinterwordspacing
S.~Khairy, R.~Shaydulin, L.~Cincio, Y.~Alexeev, and P.~Balaprakash,
  ``Reinforcement-learning-based variational quantum circuits optimization for
  combinatorial problems.'' [Online]. Available:
  \url{https://arxiv.org/abs/1911.04574}
\BIBentrySTDinterwordspacing

\bibitem{Bayerstadler.2021}
A.~Bayerstadler, G.~Becquin, J.~Binder, T.~Botter, H.~Ehm, T.~Ehmer,
  M.~Erdmann, N.~Gaus, P.~Harbach, M.~Hess, J.~Klepsch, M.~Leib, S.~Luber,
  A.~Luckow, M.~Mansky, W.~Mauerer, F.~Neukart, C.~Niedermeier, L.~Palackal,
  R.~Pfeiffer, C.~Polenz, J.~Sepulveda, T.~Sievers, B.~Standen, M.~Streif,
  T.~Strohm, C.~Utschig-Utschig, D.~Volz, H.~Weiss, and F.~Winter, ``Industry
  quantum computing applications,'' \emph{EPJ Quantum Technology}, vol.~8,
  no.~1, p.~25, 2021.

\bibitem{Lucas.2014}
A.~Lucas, ``Ising formulations of many {NP} problems,'' \emph{Frontiers in
  Physics}, vol.~2, p.~5, 2014.

\bibitem{Martonak.2004}
R.~Marton{\'a}k, G.~E. Santoro, and E.~Tosatti, ``Quantum annealing of the
  traveling-salesman problem,'' \emph{Physical Review E}, vol.~70, no.~5, p.
  057701, 2004.

\bibitem{Hogg.2003}
T.~Hogg, ``Adiabatic quantum computing for random satisfiability problems,''
  \emph{Physical Review A}, vol.~67, no.~2, p. 022314, 2003.

\bibitem{Gabor.2019}
T.~Gabor, S.~Zielinski, S.~Feld, C.~Roch, C.~Seidel, F.~Neukart, I.~Galter,
  W.~Mauerer, and C.~Linnhoff-Popien, ``Assessing solution quality of {3SAT} on
  a quantum annealing platform,'' in \emph{Quantum Technology and Optimization
  Problems}, S.~Feld and C.~Linnhoff-Popien, Eds.\hskip 1em plus 0.5em minus
  0.4em\relax {Springer International Publishing}, 2019, pp. 23--35.

\bibitem{portfolioOpt}
\BIBentryALTinterwordspacing
P.~Pierre-Louis, H.~Tong, and J.~McFarland, ``Portfolio optimization,'' 2021.
  [Online]. Available:
  \url{https://github.com/dwave-examples/portfolio-optimization}
\BIBentrySTDinterwordspacing

\bibitem{kruger2020quantum}
T.~Kr{\"u}ger and W.~Mauerer, ``Quantum annealing-based software components: An
  experimental case study with {SAT} solving,'' in \emph{Proceedings of the
  IEEE/ACM 42nd International Conference on Software Engineering Workshops},
  2020, pp. 445--450.

\bibitem{sax2020approximate}
I.~Sax, S.~Feld, S.~Zielinski, T.~Gabor, C.~Linnhoff-Popien, and W.~Mauerer,
  ``Approximate approximation on a quantum annealer,'' in \emph{Proceedings of
  the 17th ACM International Conference on Computing Frontiers}, 2020, pp.
  108--117.

\bibitem{antennaSel}
\BIBentryALTinterwordspacing
V.~Goliber, H.~Tong, and R.~Stevanovic, ``Antennas selection,'' 2021. [Online].
  Available: \url{https://github.com/dwave-examples/antenna-selection}
\BIBentrySTDinterwordspacing

\bibitem{maxCut}
\BIBentryALTinterwordspacing
------, ``Maximum cut,'' 2021. [Online]. Available:
  \url{https://github.com/dwave-examples/maximum-cut}
\BIBentrySTDinterwordspacing

\bibitem{OwenLockwood.2020_GH}
\BIBentryALTinterwordspacing
{Owen Lockwood} and {Mei Si}, ``Reinforcement learning with quantum variational
  circuit,'' 2020. [Online]. Available:
  \url{https://github.com/lockwo/quantum_computation}
\BIBentrySTDinterwordspacing

\bibitem{Skolik_GH}
\BIBentryALTinterwordspacing
A.~Skolik, S.~Jerbi, and V.~Dunjko, ``Quantum agents in the gym: a variational
  quantum algorithm for deep q-learning.'' [Online]. Available:
  \url{https://github.com/askolik/quantum_agents}
\BIBentrySTDinterwordspacing

\bibitem{Chen.2020_GH}
\BIBentryALTinterwordspacing
S.~Y.-C. Chen, C.-H.~H. Yang, J.~Qi, P.-Y. Chen, X.~Ma, and H.-S. Goan,
  ``Variational quantum circuits for deep reinforcement learning,'' 2021.
  [Online]. Available:
  \url{https://github.com/ycchen1989/Var-QuantumCircuits-DeepRL}
\BIBentrySTDinterwordspacing

\bibitem{mitarai2018quantum}
K.~Mitarai, M.~Negoro, M.~Kitagawa, and K.~Fujii, ``Quantum circuit learning,''
  \emph{Physical Review A}, vol.~98, no.~3, p. 032309, 2018.

\bibitem{hellstern2021analysis}
G.~Hellstern, ``Analysis of a hybrid quantum network for classification
  tasks,'' \emph{IET Quantum Communication}, 2021.

\bibitem{farhi2018classification}
\BIBentryALTinterwordspacing
E.~Farhi and H.~Neven, ``Classification with quantum neural networks on near
  term processors,'' 2018. [Online]. Available:
  \url{https://arxiv.org/abs/1802.06002}
\BIBentrySTDinterwordspacing

\bibitem{barison2020quantum}
\BIBentryALTinterwordspacing
S.~Barison, D.~E. Galli, and M.~Motta, ``Quantum simulations of molecular
  systems with intrinsic atomic orbitals,'' 2020. [Online]. Available:
  \url{https://arxiv.org/abs/2011.08137}
\BIBentrySTDinterwordspacing

\bibitem{barison2020quantum_GH}
\BIBentryALTinterwordspacing
------, ``Quantum simulations of molecular systems with intrinsic atomic
  orbitals,'' 2020. [Online]. Available:
  \url{https://github.com/StefanoBarison/quantum_simulation_with_IAO}
\BIBentrySTDinterwordspacing

\bibitem{copenhaver2021using_GH}
\BIBentryALTinterwordspacing
J.~Copenhaver, A.~Wasserman, and B.~Wehefritz-Kaufmann, ``Using quantum
  annealers to calculate ground state properties of molecules,'' 2021.
  [Online]. Available:
  \url{https://github.com/jcopenh/Quantum-Chemistry-with-Annealers}
\BIBentrySTDinterwordspacing

\bibitem{copenhaver2021using}
------, ``Using quantum annealers to calculate ground state properties of
  molecules,'' \emph{The Journal of Chemical Physics}, vol. 154, no.~3, p.
  034105, 2021.

\bibitem{Mauerer:2022}
\BIBentryALTinterwordspacing
W.~Mauerer and S.~Scherzinger, ``1-2-3 reproducibility for quantum software
  experiments,'' 2022. [Online]. Available:
  \url{https://arxiv.org/abs/2201.12031}
\BIBentrySTDinterwordspacing

\bibitem{ramsauer:2019}
R.~Ramsauer, D.~Lohmann, and W.~Mauerer, ``The list is the process: Reliable
  pre-integration tracking of commits on mailing lists,'' in \emph{Proceedings
  of the 41st International Conference on Software Engineering (ICSE '19)}, May
  2019, pp. 807--818.

\bibitem{Zhao.2021}
{Jianjun Zhao}, ``Some size and structure metrics for quantum software,'' in
  \emph{2021 IEEE/ACM 2nd International Workshop on Quantum Software
  Engineering (Q-SE)}, 2021, pp. 22--27.

\bibitem{mamun.2017}
M.~A.~A. Mamun, C.~Berger, and J.~Hansson, ``Correlations of software code
  metrics: An empirical study,'' in \emph{Proceedings of the 27th International
  Workshop on Software Measurement and 12th International Conference on
  Software Process and Product Measurement}.\hskip 1em plus 0.5em minus
  0.4em\relax ACM, 2017, p. 255–266.

\bibitem{schuld2019evaluating}
M.~Schuld, V.~Bergholm, C.~Gogolin, J.~Izaac, and N.~Killoran, ``Evaluating
  analytic gradients on quantum hardware,'' \emph{Physical Review A}, vol.~99,
  no.~3, p. 032331, 2019.

\bibitem{Bellman1957}
R.~Bellman, ``A markovian decision process,'' \emph{Indiana Univ. Math. J.},
  vol.~6, pp. 679--684, 1957.

\bibitem{sutton2018}
R.~S. Sutton and A.~G. Barto, \emph{Reinforcement Learning: An
  Introduction}.\hskip 1em plus 0.5em minus 0.4em\relax A Bradford Book, 2018.

\bibitem{WatkinsDayan92}
C.~J. C.~H. Watkins and P.~Dayan, ``Q-learning,'' \emph{Machine Learning},
  vol.~8, no.~3, pp. 279--292, May 1992.

\bibitem{mnih2013playing}
\BIBentryALTinterwordspacing
V.~Mnih, K.~Kavukcuoglu, D.~Silver, A.~Graves, I.~Antonoglou, D.~Wierstra, and
  M.~Riedmiller, ``Playing {Atari} with deep reinforcement learning,'' 2013.
  [Online]. Available: \url{https://arxiv.org/abs/1312.5602}
\BIBentrySTDinterwordspacing

\bibitem{van2016deep}
H.~Van~Hasselt, A.~Guez, and D.~Silver, ``Deep reinforcement learning with
  {Double Q-learning},'' in \emph{Proceedings of the AAAI conference on
  artificial intelligence}, vol.~30, no.~1, 2016.

\bibitem{tensorflow_developers_2021_5645375}
\BIBentryALTinterwordspacing
T.~Developers, ``Tensorflow,'' Nov. 2021. [Online]. Available:
  \url{https://doi.org/10.5281/zenodo.5645375}
\BIBentrySTDinterwordspacing

\bibitem{pytorch}
\BIBentryALTinterwordspacing
A.~Paszke, S.~Gross, F.~Massa, A.~Lerer, J.~Bradbury, G.~Chanan, T.~Killeen,
  Z.~Lin, N.~Gimelshein, L.~Antiga, A.~Desmaison, A.~Kopf, E.~Yang, Z.~DeVito,
  M.~Raison, A.~Tejani, S.~Chilamkurthy, B.~Steiner, L.~Fang, J.~Bai, and
  S.~Chintala, ``Pytorch: An imperative style, high-performance deep learning
  library,'' in \emph{Advances in Neural Information Processing Systems 32},
  H.~Wallach, H.~Larochelle, A.~Beygelzimer, F.~d\textquotesingle
  Alch\'{e}-Buc, E.~Fox, and R.~Garnett, Eds.\hskip 1em plus 0.5em minus
  0.4em\relax Curran Associates, Inc., 2019, pp. 8024--8035. [Online].
  Available:
  \url{http://papers.neurips.cc/paper/9015-pytorch-an-imperative-style-high-performance-deep-learning-library.pdf}
\BIBentrySTDinterwordspacing

\bibitem{DBLP:reference/db/Roy018}
\BIBentryALTinterwordspacing
P.~Roy and S.~Sudarshan, ``Multi-query optimization,'' in \emph{Encyclopedia of
  Database Systems, Second Edition}, L.~Liu and M.~T. {\"{O}}zsu, Eds.\hskip
  1em plus 0.5em minus 0.4em\relax Springer, 2018. [Online]. Available:
  \url{https://doi.org/10.1007/978-1-4614-8265-9_239}
\BIBentrySTDinterwordspacing

\bibitem{Bian.2010}
Z.~Bian, F.~Chudak, W.~Macready, and G.~Rose, ``The {Ising} model: Teaching an
  old problem new tricks,'' D-Wave Systems Inc, Tech. Rep., 2010.

\bibitem{McClean.2016}
J.~R. McClean, J.~Romero, R.~Babbush, and A.~Aspuru-Guzik, ``The theory of
  variational hybrid quantum-classical algorithms,'' \emph{New Journal of
  Physics}, vol.~18, no.~2, p. 023023, Feb. 2016.

\bibitem{Farhi.2014}
\BIBentryALTinterwordspacing
E.~Farhi, J.~Goldstone, and S.~Gutmann, ``A quantum approximate optimization
  algorithm,'' 2014. [Online]. Available: \url{https://arxiv.org/abs/1411.4028}
\BIBentrySTDinterwordspacing

\bibitem{schoenberger2021quantumqueryopt}
M.~Schönberger, ``Applicability of quantum computing on database query
  optimization,'' Master's thesis, 2021, unpublished.

\bibitem{fankhauser2021multiple}
\BIBentryALTinterwordspacing
T.~Fankhauser, M.~E. Solèr, R.~M. Füchslin, and K.~Stockinger, ``Multiple
  query optimization using a hybrid approach of classical and quantum
  computing,'' 2021. [Online]. Available:
  \url{https://arxiv.org/abs/2107.10508}
\BIBentrySTDinterwordspacing

\bibitem{docplex}
\BIBentryALTinterwordspacing
{IBM}, ``{IBM} decision optimization {CPLEX} modeling for python,'' 2021.
  [Online]. Available:
  \url{https://ibmdecisionoptimization.github.io/docplex-doc/}
\BIBentrySTDinterwordspacing

\bibitem{Bass:2012}
L.~Bass, P.~Clements, and R.~Kazman, \emph{Software Architecture in Practice},
  3rd~ed.\hskip 1em plus 0.5em minus 0.4em\relax Addison-Wesley Professional,
  2012.

\bibitem{Grover.1996}
L.~K. Grover, ``A fast quantum mechanical algorithm for database search,'' in
  \emph{Proceedings of the twenty-eighth annual ACM symposium on Theory of
  computing - STOC '96}.\hskip 1em plus 0.5em minus 0.4em\relax ACM, 1996, pp.
  212--219.

\end{thebibliography}

\end{document}